\documentclass[sigconf]{acmart}
\def\BibTeX{{\rm B\kern-.05em{\sc i\kern-.025em b}\kern-.08emT\kern-.1667em\lower.7ex\hbox{E}\kern-.125emX}}
    
\usepackage{nicefrac}
\usepackage{siunitx}
\usepackage{array,framed}
\usepackage{booktabs}
\usepackage{
  color,
  float,
  epsfig,
  wrapfig,
  graphics,
  graphicx,
  subcaption
}
\usepackage{textcomp}
\usepackage{setspace}
\usepackage{latexsym,fancyhdr,url}
\usepackage{enumerate}
\usepackage{algorithm2e}
\usepackage{algpseudocode}
\usepackage{graphics}
\usepackage{xparse} 
\usepackage{xspace}
\usepackage{multirow}
\usepackage{csvsimple}
\usepackage{balance}

\usepackage{
  tikz,
  pgfplots,
  pgfplotstable
}
\usepackage{hyperref}

\usetikzlibrary{
  shapes.geometric,
  arrows,
  external,
  pgfplots.groupplots,
  matrix
}

\pgfplotsset{compat=1.9}

\usepackage{mathtools,}

\newcommand{\chalperiod}{T_{cp}}
\DeclareMathAlphabet{\mathcal}{OMS}{cmsy}{m}{n}

\DeclareGraphicsExtensions{%
    .png,.PNG,%
    .pdf,.PDF,%
    .jpg,.mps,.jpeg,.jbig2,.jb2,.JPG,.JPEG,.JBIG2,.JB2}

\usepackage{xparse}
\newcommand{\bnm}{\begin{newmath}}
\newcommand{\enm}{\end{newmath}}

\newcommand{\bea}{\begin{eqnarray*}}%
\newcommand{\eea}{\end{eqnarray*}}%

\newcommand{\bne}{\begin{newequation}}
\newcommand{\ene}{\end{newequation}}

\newcommand{\bal}{\begin{newalign}}
\newcommand{\eal}{\end{newalign}}

\newenvironment{newalign}{\begin{align}%
\setlength{\abovedisplayskip}{4pt}%
\setlength{\belowdisplayskip}{4pt}%
\setlength{\abovedisplayshortskip}{6pt}%
\setlength{\belowdisplayshortskip}{6pt} }{\end{align}}

\newenvironment{newmath}{\begin{displaymath}%
\setlength{\abovedisplayskip}{4pt}%
\setlength{\belowdisplayskip}{4pt}%
\setlength{\abovedisplayshortskip}{6pt}%
\setlength{\belowdisplayshortskip}{6pt} }{\end{displaymath}}

\newenvironment{newequation}{\begin{equation}%
\setlength{\abovedisplayskip}{4pt}%
\setlength{\belowdisplayskip}{4pt}%
\setlength{\abovedisplayshortskip}{6pt}%
\setlength{\belowdisplayshortskip}{6pt} }{\end{equation}}

\newcounter{ctr}

\newcounter{mytable}
\def\mytable{\begin{centering}\refstepcounter{mytable}}
\def\endmytable{\end{centering}}

\newcounter{myfig}
\def\myfig{\begin{centering}\refstepcounter{myfig}}
\def\endmyfig{\end{centering}}

\newlength{\saveparindent}
\newlength{\saveparskip}
\newcommand{\E}{{\rm I\kern-.3em E}}

\renewcommand{\eqref}[1]{\mbox{Equation~(\ref{#1})}}

\def \part {part}

\renewcommand{\paragraph}[1]{\vspace*{6pt}\noindent\textbf{#1}\;}

\def \blackslug{\hbox{\hskip 1pt \vrule width 4pt height 8pt
    depth 1.5pt \hskip 1pt}}
\def \qed{\quad\blackslug\lower 8.5pt\null\par}

\newcounter{mynote}[section]

\newcommand\ignore[1]{}

\newcounter{rcnote}[section]

\newcounter{mrnote}[section]

\newcounter{fknote}[section]

\newcounter{anote}[section]

\DeclareMathSymbol{\mlq}{\mathord}{operators}{``}
\DeclareMathSymbol{\mrq}{\mathord}{operators}{`'}

\newcommand{\rhf}[2]{R_{f, \gamma}}

\DeclareDocumentCommand{\edist}{o o}{
  \ensuremath{
    \IfNoValueTF{#1}{{d}}{{\sf d}(#1,#2)}
  }
}

\newcommand{\olrk}[1]{\ifx\nursymbol#1\else\!\!\mskip4.5mu plus 0.5mu\left(\mskip0.5mu plus0.5mu #1\mskip1.5mu plus0.5mu \right)\fi}

\NewDocumentCommand{\indseq}{ O{1} O{r} }{{#1}\ldots {#2}}

\newcommand{\signed}[1]{\langle #1 \rangle}
\newcommand{\query}{\textsf{query}}
\newcommand{\sslash}{\textsf{slash}}
\newcommand{\lceco}{$\mathsf{LC}_{\mathsf{eco}}$\xspace}
\newcommand{\lcins}{$\mathsf{LC}_{\mathsf{ins}}$\xspace}

\newcommand{\tcov}{$T_{cov}$\xspace}
\newcommand{\vcov}{$V_{cov}$\xspace}

\NewDocumentCommand{\codeword}{v}{%
\texttt{\textcolor{blue}{#1}}%
}

\renewcommand\footnotetextcopyrightpermission[1]{} 
\acmYear{2024}
\begin{document}
\fancyhead{}
\def\thetitle{Unconditionally Safe Light Client}
\title{\thetitle}

\author{Niusha Moshrefi}
\affiliation{\small{Princeton University}}

\author{Peiyao Sheng}
\affiliation{\small{Witness Chain}}

\author{Soubhik Deb}
\affiliation{\small{EigenLabs}}

\author{Sreeram Kannan}
\affiliation{\small{EigenLabs}}

\author{Pramod Viswanath}
\affiliation{\small{Witness Chain}}
\date{}

\begin{abstract}
Blockchain applications often rely on lightweight clients to access and verify on-chain data efficiently without the need to run a resource-intensive full node. These light clients must maintain robust security to protect the blockchain's integrity for users of applications built upon it, achieving this with minimal resources and without significant latency.  Moreover, different applications have varying security needs. This work focuses on addressing these two key requirements in the context of Proof-of-Stake (PoS) blockchains and identifying the fundamental cost-latency trade-offs to achieve tailored, optimal security for each light client. 

The key security guarantee of PoS blockchains is {\em economic} (implied by the ``stake"). In this paper we formalize this {\em cryptoeconomic} security to light clients, ensuring that the cost of corrupting the data provided to light clients must outweigh the potential profit, thereby economically deterring malicious actors. We further introduce ``insured" cryptoeconomic security to light clients, providing {\em unconditional} protection via the attribution of adversarial actions and the consequent slashing of stakes. The divisible and fungible nature of stake facilitates {\em programmable} security, allowing for customization of the security level and insurance amount according to the specific needs of different applications.

We implemented the protocols in less than 1000 lines of Solidity and TypeScript code~\cite{code-repo} and evaluated their gas cost, latency, and the computational overhead. 
For example, for a transaction with value of \$32k, the light client can choose between zero cost with a latency of 5 hours or instant confirmation with an insurance cost of \$7.45. Thus, the client can select the optimal point on the latency-cost trade-off spectrum that best aligns with its needs. Light clients require negligible storage and face minimal computational costs, typically verifying only a few signatures (as few as one in most cases). 

\end{abstract}

\maketitle
\pagestyle{plain}

\section{Introduction}
\label{sec:intro}

In PoS blockchains, validators secure the network by ``locking"  a certain amount of stake (e.g., 32 ETH in Ethereum) to participate in the consensus protocols. The inherent nature of security in these PoS blockchains is {\em economic}: the greater the total stake, the more cost or loss needed to attack the consensus protocol (e.g., one third of the total stake, as in Ethereum PoS). Enforcing such costs relies on a feature called {\em accountable}  security \cite{Sheng_2021,cryptoeprint:2021/628,civit2019polygraph,buterin2017casper,tang2023raft,buchman2016tendermint}, which allows for the confiscating or ``slashing'' of stakes if validators sign conflicting states. 

\textit{Full nodes} in PoS blockchains play a critical role in maintaining the blockchain's integrity. They verify consensus signatures, replicate a full copy of transaction history, and execute state transitions. These tasks require significant resources and sophisticated hardware. For example, operating a full node for Ethereum demands at least 2TB of SSD storage~\cite{Ethereum2023RunNode}. 
In contrast, light clients prioritize resource efficiency, making them suitable for applications that only need to verify specific transactions and states, such as mobile wallets and cross-chain bridges~\cite{goes2020interblockchain,mobilewallet}. Due to their limited resources, light clients sacrifice some degree of independence and immediacy in verifying blockchain security. They must communicate with full nodes to achieve the same level of security, particularly across three main areas: consensus (agreement on data inclusion), data availability (preventing censorship and downtime) and the validity of state transitions (ensuring state consistency). This paper explores the fundamental trade-offs between cost and latency of light clients required to achieve optimal security. We specifically focus on light clients that verify the consensus agreement and isolate the problem from state transition validation and data availability check, which are topics of independent interest~\cite{yu2020coded,albassam2019fraud,sheng2021aced,kalodner2018arbitrum}.

\begin{table*}[ht]
\centering
\caption{Comparison of our designs \lceco and \lcins with existing light client protocols. Computation refers to the number of signatures required to be verified. For full PoS and sync committee, we assume all states are synchronized and only calculate the cost of verifying the consensus of a single block. Storage here is only for the consensus and not for storing the entire blockchain. For \lceco, the latency is determined by the light client and typically spans a few hours. The distribution of Ethereum transaction sizes indicates that over 85\% of transactions are valued at less than 10 ETH~\cite{ethereum-analysis}, requiring only one data provider with a 32 ETH stake to secure them. Therefore, we focus our calculations on scenarios that rely on a single data provider. $V$ is the total number of validators of PoS blockchain ($V$ is greater than 1 million as of March 2024 for Ethereum\cite{validatorcount-ethereum}).}
    \begin{tabular}{|c|c|c|c|c|c|}
        \hline
        \textbf{-} & \textbf{Full PoS} & \textbf{Sync Committee} & \lceco & \lcins \\ \hline
        \textbf{Computation} & $V/$32 & 512 & 1 & 1 \\ \hline 
        \textbf{Storage} & 100MB & 30KB & <100B & <100B \\ \hline 
        \textbf{Latency} & 10 secs & 1 sec & 5 hours & 2.8 ms\\ \hline
        \textbf{Security} & Economic & Reputation-based & Economic & Insured \\ \hline
        \textbf{Programmability} & $\times$ & $\times$ & \checkmark & \checkmark \\ \hline
    \end{tabular}
\end{table*}
    
Bitcoin introduces Simple Payment Verification (SPV) as its light client protocol. SPV enables light clients to verify the inclusion of a transaction in a specific block using a Merkle proof and the block header. Therefore, light clients need only download the block headers of the blockchain and can verify transaction finality by checking the depth of the block. In this context, the computational cost of consensus verification is relatively low for light clients in Bitcoin. However, in PoS blockchains like Ethereum, the consensus check is inherently more complex by design. It involves maintaining the whole validator set, tracking their stake changes and performing many signature checks for the consensus protocol. On the other hand, the security of PoW light clients relies on the assumption that the majority of full nodes are honest. In contrast, PoS blockchains derive their security economically through slashable stakes. The system rely on the rationality of consensus participants, aiming to ensure that the cost of attack exceeds any potential profit, which may vary in different contexts. Therefore, designing a light client protocol for PoS blockchains presents two essential requirements: (1) addressing the high cost of consensus verification, and (2) ensuring economic security and leveraging its dynamic nature.

To reduce the cost of verification, Ethereum's current light client protocol relies on a sync committee~\cite{sync-committee}, composed of 512 randomly selected Ethereum validators, each staking 32 non-slashable ETH. However, this design exhibits significant security flaws. A dishonest supermajority within the sync committee could mislead light clients to accept invalid data without facing any penalties. Even if accountability were introduced through slashing, the combined stake of the sync committee remains negligible in comparison to the extensive Ethereum validator pool, which exceeds over 1 million validators as of March 2024~\cite{validatorcount-ethereum}. Consequently, this approach provides weak security for the light clients.

Furthermore, in the current design, light clients treat all transactions, whether worth a million dollars or a hundred dollars, with the same level of security. However, in the real world, security measures are naturally tailored based on the value at risk. For example, banks allocate more resources and scrutiny to safeguarding substantial deposits than they do for smaller checks. This principle should ideally extend to blockchain transactions, where the security guarantee provided by a light client should correspond to the transaction's value. For instance, if a light client is verifying the inclusion of a \$100 transaction, it should require authentication supported by an amount of slashable stakes slightly higher than \$100. In such a scenario, a rational validator staking this amount would lack the incentive to manipulate the data sent to that particular light client, as the potential loss from being slashed would outweigh the gain from the fraud.



In this paper, we decouple the security of the blockchain consensus from the security provided to light clients for accessing on-chain data. For light clients, there is no need to comprehensively verify the consensus of the entire network. Instead, we apply the same principles of \textit{economic security} for PoS system to each light client and introduce \textit{programmability} into this security framework: we ensure that the cost of corrupting a light client's verification process exceeds the potential profit from such corruption. This cost is tailored to the specific security needs of light clients, akin to the k-deep confirmation rule in Bitcoin. Consequently, each light client can independently balance its security level against the associated verification cost.


With security now programmable, individualized, and inherently economic, we introduce \textit{insured security}. Economic security already guarantees that the stakes slashed from malicious validators always exceeds the potential gains. Thus, even if validators behave irrationally, the economic penalties are sufficient to cover any losses incurred. Insured security is designed to provide additional financial protection in the event of security breaches. Before processing a transaction, light clients are able to purchase insurance corresponding to the transaction's value. Validators involved in the light client protocol are held accountable for their commitments: if they sign incorrect data, they are penalized by having their stakes slashed, and the insured amount is then refunded to the light client from these funds.

\begin{figure}
    \centering
    \includegraphics[width=\linewidth]{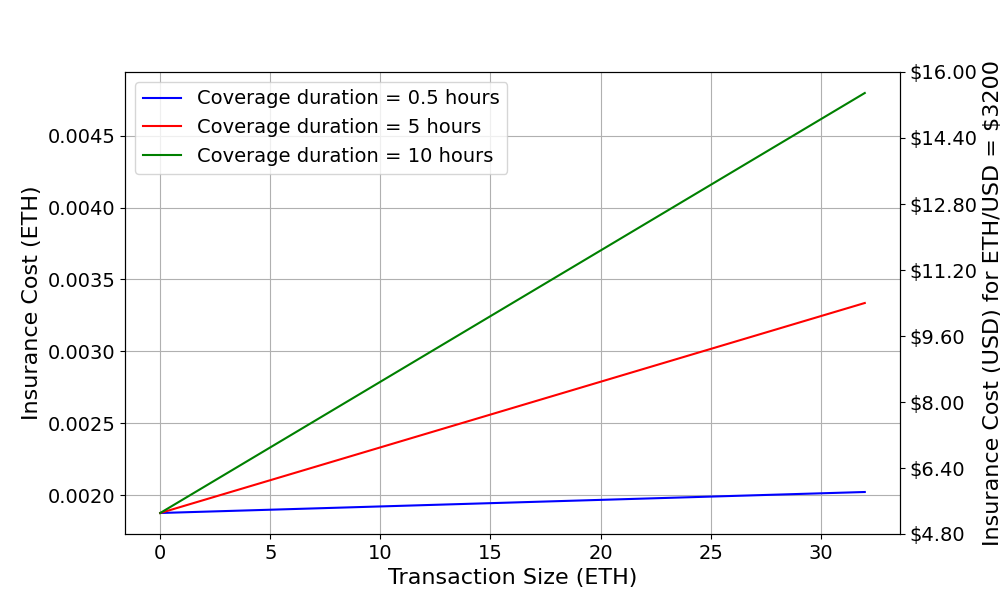}
    \caption{Insurance cost vs. Transaction size (or covered value) for different insurance duration}
    \label{fig:insurance-cost}
\end{figure}

\paragraph{Our contribution.} In this paper, we propose a light client protocol for PoS blockchains, denoted as \lceco, featuring programmable economic security and optimal cost-efficiency. Its variant \lcins incorporates an insurance scheme to further provide unconditional protection for adversarial actions. The system offers the following advancements:

\begin{itemize}
    \item \textbf{Economic security.} We define economic security for light clients by making corruption economically infeasible for validators. In this protocol, light clients who want to verify specific on-chain states interact with a group of full nodes, known as \textit{data providers}. These data providers validate and sign off on the legitimacy of the requested states, sending their confirmations to the light clients. The light clients then wait for some predetermined time to get ensured that the data providers are not slashed and the provided data is correct. During this waiting period, a network of full nodes, termed as \textit{watcher network}, actively monitors for potential inconsistency between data providers' responses and on-chain data. This network guarantees that if any data provider signs an incorrect proof, at least one watcher will detect this error and alert the light client within the designated timeframe. Through this system, the protocol effectively ensures that the costs associated with misleading behavior exceed any potential profits, thus promoting honesty through economic incentives.
    \item \textbf{On-demand, programmable security.}  Unlike traditional PoS blockchains that offer a uniform security level, our protocol allows light clients to customize security measures based on their specific application needs. Leveraging stake-based voting, the group of data providers are chosen to ensure that the cumulative stakes exceed the desired security threshold -- a minimum percentage of stake backing the data for specific applications. This approach provides each application with the granular control based on the risk assessments.
    \item \textbf{Insured security.} Our protocol further introduces an insured  security feature, enabling light clients to purchase insurance against potential losses from adversarial actions. This insurance scheme provides dual benefits: firstly, it allows light clients to accept the data from providers immediately upon receipt, bypassing the waiting period. Secondly, in the case of security attack, it ensures that light clients do not incur financial losses. The insurance cost is calculated based on the coverage duration, the value protected, and the expected return rate for insurance stakers. Additionally, a constant gas cost is incurred for the inclusion and execution of the insurance payment transaction. The cost of insurance across various coverage durations is illustrated in Figure \ref{fig:insurance-cost}. This scheme provides compensation for damages, protecting against irrational adversaries willing to incur significant penalties.
    \item \textbf{Optimal cost and latency.} Our protocol not only achieves programmable and optimal security guarantees but also significantly reduces the computational costs associated with consensus verification. Table~\ref{tab:comparison} illustrates the performance comparisons between various protocols. Our light client protocol with economic security optimizes both communication and computational efforts, although it does increase latency due to the optimistic verification path required to ensure security. By introducing insurance, the protocol achieves optimal performance across all metrics and offers enhanced financial guarantees in the event of attacks.

\end{itemize}

\paragraph{Implementation and evaluations}
We implemented our nodes in TypeScript using less than 700 lines of code, and the smart contract in Solidity using approximately 300 lines of code. Our implementation of the light client protocols is live on the Ethereum testnet Sepolia. We assume multiple data providers registered on the smart contract with 32 ETH worth of stake each, and the light client is connected to one honest watcher. Then, assess the light client design by the following metrics:
\begin{itemize}
    \item Cost efficient: Light client incurs the cost of insurance if they opt in for it. The cost is proportional to the level of desired security for the light client.
    \item Fast confirmation: The fastest a light client can get confirmation for the data is as soon as it is confirmed on Ethereum. This instant confirmation is enabled for light clients using the insurance.
    \item Light computation: The computation cost for our design is minimal; few milliseconds for a light weight device.
\end{itemize}

The light client can select an optimal balance in the trade-off between latency and cost. It has the option to either confirm the inclusion of data after a delay at no cost, or pay a cost proportional to the delay to achieve instant confirmation. For a transaction valued at most \$32k, we calculate the cost to be as In all cases, the light client maintains security equivalent to that of a full node.

\paragraph{Organization} In section \ref{sec:relwork}, we provide background on Proof-of-Stake (PoS) blockchains and light clients. Our model is introduced in section \ref{sec:model}. The protocol designs for economic security and insured security are presented in sections \ref{sec:designEco} and \ref{sec:designIns}, respectively, along with their analyses. The evaluation of the protocol and details of our implementation are discussed in section \ref{sec:eval}. Further discussion on the system design is provided in section \ref{sec:discussion}.  
\section{Background and Related Work}
\label{sec:relwork}
\subsection{Background}
\label{subsec:background}

In this section, we introduce and define key terminology essential for understanding the protocol.

Ethereum uses Proof-of-Stake (PoS) as its consensus mechanism after the upgrade in Sep 2022. PoS means that the growth of the blockchain is guaranteed by participants who put some stake at risk to be slashed in case of their misbehavior.

In the Ethereum network, \textit{validators} stake 32 ETH in a contract to be able to participate in the network, with a unique secret key and a public key serving as their identity. A validator's stake will get destroyed if it acts dishonestly. Each validator is linked to a \textit{node}. Nodes can host various validators. Currently, Ethereum has more than 800,000 validators but less than 6000 nodes. 

Ethereum operates with two crucial intervals: the \textit{slot}, 12 seconds in duration, and the \textit{epoch}, comprising 32 slots, equivalent to 6.4 minutes. One validator is randomly selected to be a block proposer in every slot. Then, other validators \textit{attest} to the proposed block based on a random selection. Every active validator attests in every epoch, but not in every slot. Dishonest attestations by a validator leads to its stake being slashed.

A transaction gets \textit{finalized} when it is part of a block that will not get out of the canonical chain without a large amount of ETH getting burned. On Ethereum, the first block in each epoch is a \textit{checkpoint}. Validators vote for pairs of checkpoints that they consider to be valid. If a pair of checkpoints attracts votes representing at least two-thirds of the total staked ETH, the later checkpoint becomes \textit{justified}, and the earlier checkpoint that was previously justified becomes \textit{finalized}.

Moreover, on Ethereum, the inclusion of a transaction can be checked by verifying an inclusion proof. A block contains several fields including a \verb+body+ which has an \verb+execution_payload+. It has 
a header called \verb+execution_payload_header+ that includes \verb+transactions_root+ using which transaction inclusion checks can be done efficiently.

Finally, \textit{light client} is a form of client that interacts with the blockchain while consuming minimal resources. They play a crucial role in enabling lightweight access to blockchain data for various applications and users.


\subsection{Related Work}
\label{subsec:relwork}

Techniques for a lightweight client to verify consensus were originally discussed in the Bitcoin paper~\cite{nakamoto}, known as Simplified Payment Verification (SPV). It allows clients to download only block headers and verify the existence of transactions through SPV proofs. While this method reduces the workload on the resource-limited client~\cite{dorri2019lsb,frey2016bringing,gervais2014privacy}, it necessitates a frequent online presence to keep up with the main chain's growth. Clients inactive for long periods face the challenge of linearly verifying block headers upon reactivation~\cite{superlightclient}. To address the limitations of SPV, innovations like FlyClient~\cite{bunz2020flyclient} and Non-Interactive Proofs of Proof-of-Work (NiPoPoW)~\cite{kiayias2020non} leverage the inherent authentication of a chain's few suffix blocks, enabling the proof of these blocks to clients at a sublinear cost. However, their reliance on verifying PoWs restricts their applicability to PoS consensus models. To address the security and efficiency challenges in PoS bootstrapping, PoPoS~\cite{agrawal2022proofs} introduces a bisection game to effectively challenge adversarial Merkle trees of PoS epochs. \cite{tas2022light} proposes a composable solution to create light clients for lazy blockchains~\cite{al2019lazyledger,bagaria2019prism}. Though they achieve minimal space and avoid requiring always-online clients to maintain stake distributions, the issue of enabling clients to go offline without incurring substantial costs for rejoining the network remains unaddressed.

In popular Ethereum wallets like MetaMask, the consensus client logic is handled by centralized infrastructure providers such as Infura~\cite{infura}. These providers undoubtedly result in a lightweight and efficient user experience. However, the centralization of such services means that a compromised provider could potentially mislead users by altering payment and balance details or by censoring transactions. To improve security and accelerate bootstrapping, one of the most popular adoptions in Ethereum is a sync committee~\cite{sync-committee}, which comprises 512 validators selected every 27 hours, to sign block headers in the beacon chain. However, the lack of economic penalties for misbehavior among committee members still raises concerns about the reliability and security of this system. The introduction of a generic superlight client~\cite{lu2020generic} for permissionless blockchains within a game-theoretic framework offers a new perspective. This model represents a special variant of multi-party computation under rational settings. However, it does not provide economic guarantees and fails to address the threat posed by malicious, irrational data providers. In contrast, our light client focuses on cryptoeconomic security. The concept of economic security paired with an insurance scheme was first explored in recent work named stakesure ~\cite{stakesure}. We apply this definition to establish a general framework that provides economic security and insured security that unconditionally protects the light client against malicious behaviors for PoS chains. 

An alternative research approach focuses on the use of zero-knowledge proofs to create succinct proofs~\cite{ben2017scalable}. For example, Mina~\cite{mina,bonneau2020coda} and Plumo~\cite{gabizon2020plumo} effectively facilitate lightweight consensus verification through the use of recursive SNARK compositions~\cite{bitansky2013recursive} and SNARK-based state transition proofs. Halo~\cite{bowe2019recursive} improves Plumo by removing the trusted setup. However, these methods impose a considerable computational burden on block producers for proof generation~\cite{chatzigiannis2022sok}, and they do not address compensation for potential losses experienced by light clients. In the context of other PoS protocols like Tendermint~\cite{buchman2016tendermint} used in Cosmos, the role of the light client is explored within their Interblockchain Communication (IBC) protocol\cite{goes2020interblockchain,braithwaite2020tendermint}. Notably, these implementations are specific to their respective platforms and are not directly applicable to Ethereum or various other PoS blockchains. 
\section{Model}
\label{sec:model}
This section introduces the model used in our protocol design.

\paragraph{Blockchain.} We assume a programmable blockchain with deterministic finalization rule for its blocks. On the Ethereum blockchain, a block is finalized after at least two subsequent epochs come on top of it, typically taking around 13 minutes.

\paragraph{Slashing smart contract.} The protocol includes an on-chain slashing contract adhering to standard smart contract abstractions. It can access the block hash of the previous blocks in the blockchain. All parties can send messages to this contract.

\begin{figure}
    \centering
    \includegraphics[width=0.45\textwidth]{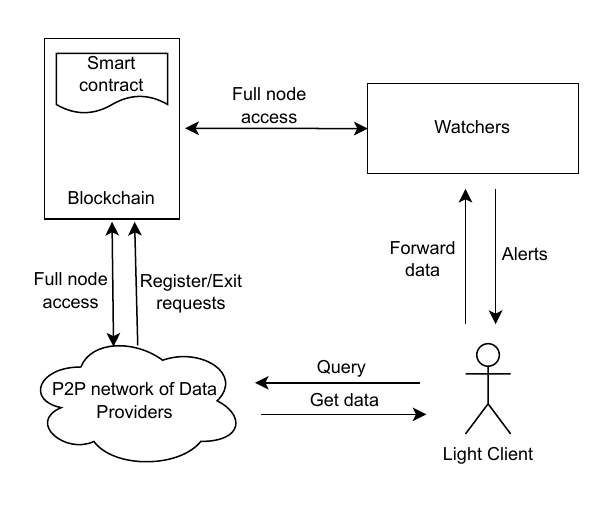}
    \caption{System participants and their interactions.}
    \label{fig:parties}
\end{figure}

The parties involved in the system are: data providers, watchers and the light client. The parties, their relation and connections are shown in Figure \ref{fig:parties}. All parties are computationally bounded to perform only polynomial-time computations. 

\paragraph{Data providers.} Data providers operate full nodes and keep track of the latest state of the blockchain. They stake assets to provide services to verify the validity of states requested by light clients. The stakes are subject to potential slashing for misbehavior to ensure accountability. Each data provider has a publicly known cryptographic identity, referred to as a public key, which is linked to their stake. They sign all data sent to light clients with the secret key corresponding to their public key, enabling verification of data origin and integrity.  We use $\signed{m}_{sk}$ to represent a message $m$ signed by the data provider with secret key $sk$.

Data providers can join and stay in the system by maintaining a minimum amount of stake. Moreover, they can freely exit, reclaiming their remaining stake from the system after a withdrawal delay.
Data providers may act arbitrarily maliciously. All malicious data providers are governed by one adversary and can coordinate attacks. We assume the existence of at least one rational data provider to ensure liveness.



\paragraph{Watchers.} 
Watchers are full nodes connected to light clients to assist in data verification. Anyone can become a watcher to profit from monitoring and slashing misbehaving parties.  Watchers receive data provided to the light clients by the data provider. When a watcher receives invalid data signed by a data provider, it is obligated to present on-chain evidence to slash the misbehaving provider's stake and alert the light client before the \textit{challenge period} ends. The challenge period (denoted by $T_{cp}$) is a duration determined by light clients based on the value of the state it wants to check. A longer challenge period increases the light client's confidence that an honest watcher has verified its provided data. The challenge period has a maximum duration $maxT_{cp}$ determined by the system when initializing the protocol. For simplicity, we assume each light client connects to at least one honest watcher. Building a watcher network within rational model is an independent topic of interest~\cite{sheng2024proof}.


\paragraph{Light client.} Light clients are resource-constrained clients that aim to verify the inclusion of a state or transaction on the blockchain with minimal cost. Light clients are connected with a group of data providers and watchers during the verification process. They can not directly read data from blockchain, but can send transactions to on-chain contracts through the network diffusion functionality.


\paragraph{Heavy checks.}
There exist a couple of ways for a client to access the latest finalized state of the blockchain. These methods are heavy in computation, but provide full node level security. For example, to use the full node protocol or zero-knowledge proofs. We use these methods during bootstrapping phase(one time event for each light client) and in the dispute path of our protocol (happening infrequently).

\paragraph{Network.} The model assumes synchronous communication for all parties, with maximum message delay $\Delta$. The data providers form a peer-to-peer (p2p) network where they propagate data using a gossip protocol. Light clients get connected to some data providers as access points to the network to send their queries and receive their response.

\section{Light Client With Economic Security}
\label{sec:designEco}


\subsection{Protocol Overview}
\label{subsec:designEco:overview}
We first introduce our light client protocol, \lceco, which enables light clients to access on-chain data with economic security. A light client queries specific data providers to verify the inclusion of a state or transaction on the blockchain, referred to throughout the paper as the \textit{target state}. The process itself is known as the \textit{state inclusion check}. Data providers sign the block hash containing the target state along with any necessary proofs, and then send this information to the light client. The light client forwards all received data to connected watchers and waits for the challenge period to pass, during which it listens for alerts from watchers. If no alerts are received, the light client verifies the signatures and the proof, accepting the validity of the data. If watchers receive data from a light client, they check it against their view of the blockchain. If found inconsistency, the watchers slash the data provider on-chain and alert the light client about the discrepancy. We will delve into the details of each component's design in the following sections.

\subsection{Light Client}
\label{subsec:designEco:lightClient}
Before checking the state, the light client performs a ``heavy check'' to bootstrap and obtain the latest set of data providers with their stakes (see details for updating process in \ref{subsec:designEco:dataProviders}). Then the light client decides on a challenge period duration (\(T_{cp}\)), which determines how long to wait while listening for alerts from watchers before accepting the state. The length of $T_cp$ depends on the estimated time it takes for an honest watcher to receive and verify the data after it is forwarded by the light client. Extending the challenge period enhances security but also increases the protocol's latency.  

After bootstrapping, the light client selects a group of data providers, ensuring the cumulative stakes exceed the value of the requested state. A higher total stake increases the light client’s confidence in the accuracy of the data. However, this also necessitates more computations to verify proofs, presenting a trade-off that the light client must carefully manage when choosing data providers.

Afterwards, the light client generates a request $\query(n_B, h_s)$ and sends it to the chosen data providers to verify that a target state $s$ has been included in a committed block $B$. This request includes the block number $n_B$ of the block containing the target state and the hash of the target state $h_s$.

\begin{figure}
    \centering
    \includegraphics[width=1\linewidth]{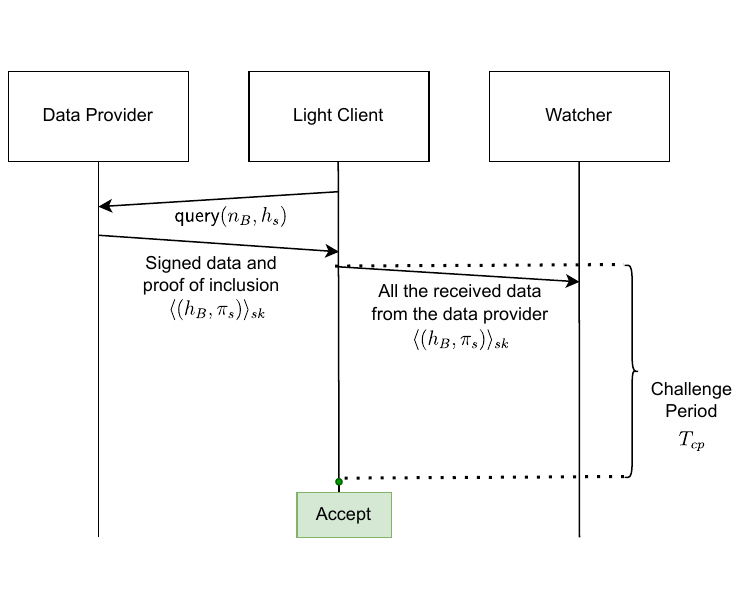}
    \caption{The normal path for economic security. We have simplified this example, by supposing the light client has chosen only one data provider. Also, we only show the honest watcher here.}
    \label{fig:eco-happy}
\end{figure}

After sending requests, the light client waits for responses from the data providers. Upon receiving these responses, it forwards the data to the connected watchers and then monitors for any alerts. If no alerts are received once the challenge period elapses, the light client verifies the signatures of the data providers and examines the proof of state inclusion. Upon successful verification, the light client deems the data trustworthy and proceeds to accept it. The flow of normal path is depicted in Figure \ref{fig:eco-happy}.


\begin{figure}
    \centering
    \includegraphics[width=1\linewidth]{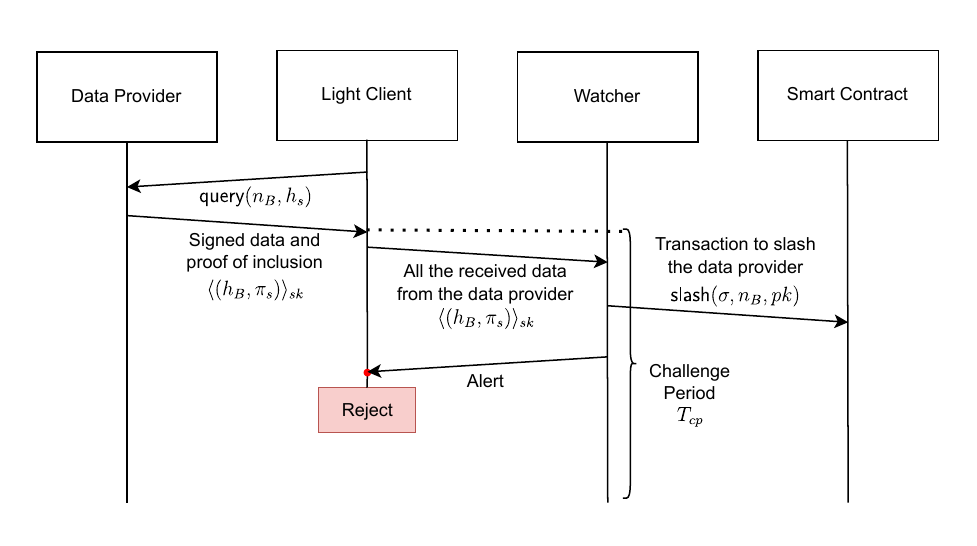}
    \caption{The dispute resolution path for economic security. This simplified example assumes the light client has selected a single data provider and depicts only the honest watcher.}
    \label{fig:eco-dispute}
\end{figure}

On the other hand, the light client may encounter the dispute path when receiving alerts from watchers. In this case, it verifies the alert, discards all data from the implicated data provider, and restarts the protocol. Alerts include an inclusion proof for the slashing event. The light client must perform a ``heavy check'' to ensure that the slashing event actually occurs. Note that this intensive verification is specific to the infrequent dispute path.
The interactions during the dispute resolution path are depicted in Figure \ref{fig:eco-dispute}.

\subsection{Data Providers}
\label{subsec:designEco:dataProviders}


Data providers respond to requests from light clients by signing the target state if it has been included in a committed block. The response is denoted as $\signed{(h_B, \pi_s)}_{sk}$, where $h_B$ represents the hash of the requested block $B$ and $\pi_s$ is the inclusion proof of state $s$ in $B$. 

\begin{figure*}
    \centering
    \includegraphics[width=16.54cm ,height=8cm]{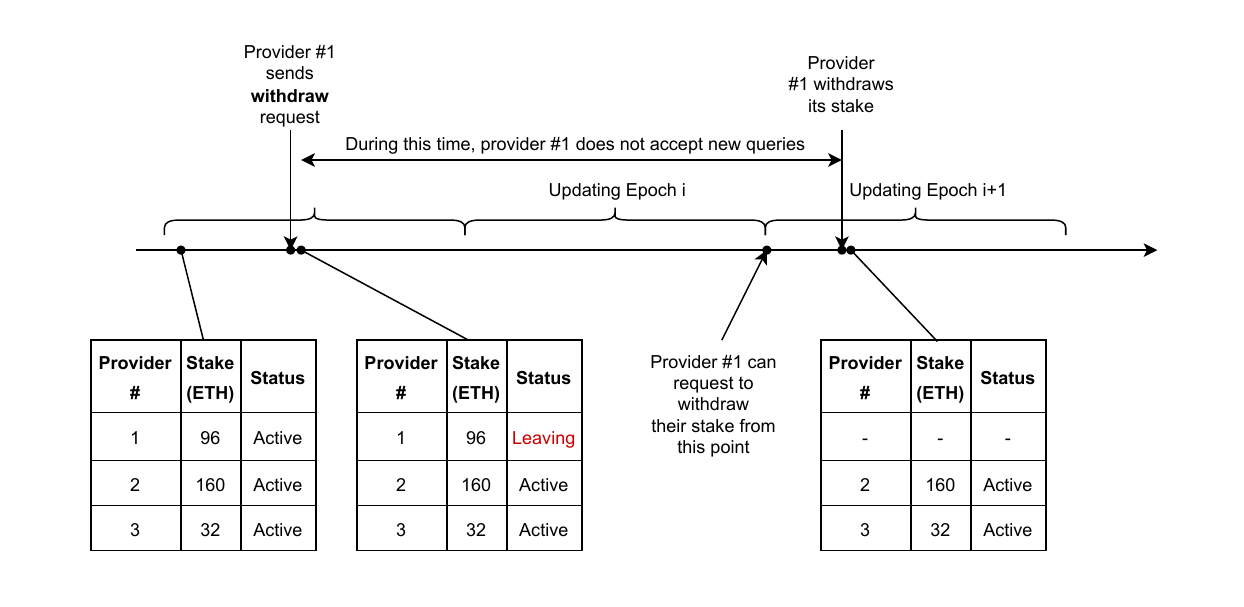}
    \caption{This diagram illustrates the blockchain's growth and details the process by which a data provider exits the protocol.}
    \label{fig:dpSetUpdate}
\end{figure*}

Data providers are free to join and leave the system. They form a set that gets updated upon execution of \textit{register} and \textit{withdraw} requests. We define \textit{update epoch} for data providers, each consisting of $B_u$ blocks, a value set during protocol initialization. The duration of update epoch, denoted by $T_u$ ($T_u = B_u \times \text{avg(block interval time)}$ represents the minimum duration needed to process withdraw requests. We ensure $T_u$ is significantly greater than the maximum challenge period ($T_u \gg maxT_{cp}$). Withdraw requests made within an update epoch take effect at the last block of the following epoch. Below, we detail how each type of request is executed:
\begin{enumerate}
    \item \textbf{Register:}  A new data provider who deposits more than the minimum required stake is immediately added to the \textit{active} data provider list. 

    \item \textbf{Withdraw:} When a data provider decides to exit the system, their status immediately changes from \textit{active} to \textit{leaving}, and they cease protocol engagement. The actual withdrawal of their stake is only permitted after the end of the following update epoch. An example of this process is illustrated in Figure \ref{fig:dpSetUpdate}.
\end{enumerate}
A data provider labeled as leaving may still be slashed for past behavior but will no longer accept requests. This measure prevents data providers from acting maliciously and quickly withdrawing their stake to avoid penalties. The condition $T_u \gg maxT_{cp}$ ensures that all user challenge periods have expired for any requests made to a data provider before they are permitted to withdraw their stakes.

After bootstrapping, the light client acquires an initial list of data providers. To stay updated, it must refresh this list each time it validates a block in a new update epoch, since the set of data providers remains static throughout an update epoch. Lengthening the update epoch delays stake withdrawal for stakers, impacting those wishing to exit the system. However, this delay occurs only once for data providers who choose to leave. The advantage of extending update epochs is a reduction in the frequency of set transitions, which simplifies the verification process for light clients. Additionally, this updating mechanism enables light clients to predict the active data provider set for future blocks, ensuring they experience no delays due to changes in the network of data providers.

A light client that remains online to perform multiple state inclusion checks does not need to bootstrap for each verification. Assume the blockchain is in update epoch \(i\), where \(i \in \mathbb{N}\). If the light client has recently completed an inclusion check, it already has access to the active data provider set for epoch \(i\). To prepare for epoch \(i+1\), it needs only the register and withdraw requests from epoch \(i-1\) to be applied to the set from epoch \(i\). The light client uses these requests as the target state in our protocol and queries data providers to confirm their inclusion on the blockchain. Once the maximum challenge period from the start of epoch \(i\) elapses, all checks for epoch \(i-1\) are considered verified from the light client's perspective. Thus, the light client can anticipate the status of the data provider set for epoch \(i+1\) before the blockchain progresses to that point, avoiding delays related to data providers entering or exiting the network, and allowing it to select active data providers in advance.

\subsection{Watchers}
\label{subsec:designEco:watchers}
Watchers receive signed data from light clients, originally provided by data providers. They continuously verify that the signatures are from active data providers and confirm whether the data has been finalized on the blockchain.

If a signature fails to meet these criteria, the watcher sends a transaction $\sslash(\sigma, n_B, pk)$ to the smart contract, providing the data provider's public key and the signature on the disputed data, along with the block number. If the data is proven incorrect or lacks finalization confirmation, the smart contract slashes the stakes of the offending data provider.

If the data does not exist on the finalized chain, it may be due to pending finalization or because another block for that number has been finalized. The former can be confirmed via the blockchain's finalization rules, while the latter requires proving to the contract that a different block has been finalized by the Ethereum consensus. Under the assumption that Ethereum validators are honest and do not endorse conflicting blocks, such proof confirms malicious actions by the data provider, warranting a slash.

Following a successful slash, the watcher notifies the light client, including proof of the slashing event's inclusion on the blockchain.

Note that if the signature does not belong to any of the current active data providers with stake at risk, the watcher cannot perform an on-chain slashing. However, the watcher will promptly notify the light client of this situation. This occurs if, after the light client selects an active data provider to query and before the challenge period concludes, the data provider's entire stake is slashed for another query, rendering them inactive. So, the watcher has to alert the light client not to trust the data this data provider provides. In such cases, the watcher is required to provide the light client with proof of the recent slashing event that deactivated the data provider. The light client then verifies the inclusion of this slashing event before disregarding the data.

\subsection{On-chain Smart Contract}
\label{subsec:designEco:contract}
An on-chain smart contract performs two key functions: first, it holds the stakes of data providers and decides when to slash them; second, it manages the entry and exit of data providers within the system. The smart contract maintains a list of public keys for all data providers, determining \textit{active} and \textit{leaving} data providers, along with their corresponding stakes.

The slashing conditions are implemented in the contract to enable on-chain verification of disputes raised by watchers, thus eliminating the need for trust in the watchers. Anyone can present proof to dispute a data provider, and the function is open to be called from arbitrary addresses. 

To prove that a signed block hash does not correspond to a finalized block, a watcher submits a conflicting finalized block hash to the contract. The smart contract verifies if this conflicting block hash is indeed finalized by checking its block number and matching it with the recorded block hash on the blockchain. Additionally, the contract assesses the finality rule of the blockchain for the block. With this information, it can confirm the validity of the watcher's dispute and, if valid, slash the offending data provider.

\subsection{Analysis}
\label{subsec:designEco:analysis}
\textbf{Safety.} 
After receiving data from the data provider, the light client waits for the challenge period to see if it receives any alerts from the watchers. This guards against potential malicious behavior by the data provider who might manipulate data despite the risk of being slashed. If no alerts are received, the light client can trust the data. However, if any alerts are received and verified, the light client discards the data and restarts the process. The light client is secure against both rational and irrational data providers because data providers are economically disincentivized from deviating from the protocol, and watchers are in place to alert the light clients about any malicious data. 


Note that our protocol aligns the light client with the state derived from the Ethereum consensus mechanism, but it does not safeguard against attacks targeting the consensus process itself. To counter potential consensus attacks on Ethereum that might affect the light client, an additional layer of verification through social consensus can be implemented.

\textbf{Full node level security.} 
Our approach offers light clients the same level of security guarantees that Ethereum PoS provides to full nodes. On Ethereum, validators risk their stakes when they vote on proposed blocks. Misbehaving validators face slashing penalties. After a certain period, a block achieves finality, at which point full nodes accept it permanently. Similarly, in our light client protocol, data providers with stakes at risk can also be slashed for misbehavior. They sign the blockchain data they assert to be finalized. Once the challenge period ($T_{cp}$) has elapsed, the data is considered finalized from the light clients' perspective and is accepted permanently. In both scenarios, economic security is achieved.

\textbf{Liveness.} 
The system remains live as long as there is at least one rational data provider in the network. In this context, "liveness" means that no query goes unanswered indefinitely. Data providers are motivated to participate and act honestly within the protocol because their profits are proportional to their stake and the services they provide to light clients. This incentivization ensures that rational data providers stay engaged and responsive.

The rate at which light clients receive responses is at least equal to the total throughput capacity of these rational data providers. Consequently, when only a few rational data providers are present, some light clients may experience delays in receiving responses. However, since the total throughput of rational data providers is non-zero, all light client requests will eventually be processed, thus ensuring the system's liveness.


\textbf{Efficiency.}
We anticipate that Ethereum validators will be the most willing parties to join as data provider nodes by re-staking. Since they already need to run a full node, this protocol incurs very minimal overhead in return for additional yield on top of what they earn as Ethereum validators. Each validator stakes 32 ETH in the Ethereum PoS. However, nodes running multiple validators can re-stake in the light client protocol as one data provider, using a single public key but providing the sum of the stakes for all the validators they run.

Many small transactions that collectively sum up to 32 ETH can be handled at the same time by a data provider re-staking as a single validator. Moreover, all those light clients only need to verify a single signature, requiring minimal computational power and allowing for quick processing.

Even larger transactions could be verified by a light client by only checking a few signatures. It depends on how many data providers can cover the value of the transaction by putting at risk their whole staked values. The number of required signature verifications increases when the light client wants to check transactions with a very high value and needs to use data from multiple data providers.

This programmable security aligns with what we observe in the real world, where many small transactions happening in banks or institutions do not need extra care and can proceed quite quickly. However, to move large amounts, additional time and care are required.

\textbf{Challenge period.}
From section ~\ref{sec:model}, our security model assumes the existence of at least one honest watcher who, during the challenge period $\chalperiod$, will examine the data forwarded by the light client. It's important to note that $\chalperiod$ is not a global parameter; instead, ; instead, each light client sets their own  $\chalperiod$ based on their specific security requirements. The longer the $\chalperiod$ that a light client considers, the higher the likelihood that an honest watcher will review the data received from the data provider(s). Beyond a specific duration, increasing the challenge time no longer change the probability of an honest watcher verifying the data significantly. We incorporate this threshold in the protocol by setting a maximum value $maxT_{cp}$.


\section{Light Client With Insured Security}
\label{sec:designIns}

In this section, we enhance the protocol to better serve light clients who require faster confirmation. To eliminate the waiting time associated with the challenge period and to compensate affected light clients connected to a malicious data provider, we introduce a new concept: \textit{insured security}. This approach assigns a portion of the data providers’ staked value to each query from a light client in the form of insurance. Then, should misbehavior occur, this insurance mechanism guarantees that the light client is compensated from the slashed staked value. This adaptation leads to the design of an insured light client protocol, \lcins, that offers unconditional safety. We will elaborate on this protocol in this section.

\subsection{Protocol Overview}
\label{subsec:designIns:overview}
The light client first calculates the maximum potential loss from being misinformed about the state of the blockchain. It then purchases insurance for the determined amount and specifies a coverage duration. The light client initiates a transaction to buy the insurance and utilizes \lceco to check its finalization. 

Next, the light client sets a local challenge period. The insurance coverage duration can accommodate multiple inclusion checks. For \textit{independent state inclusion checks}, where the challenge periods do not overlap, the insured amount only needs to cover the maximum value of these checks. Conversely, if multiple inclusion checks overlap within the same challenge period, the insured amount should cover the total value of all checks.

In this design, stakes are specifically attributed to state check queries from light clients. This ensures that the stakes are not overloaded, and overloading refers to using a portion of the stake to back multiple inclusion checks that together exceed the staked value. This specific attribution allows for compensation to the light client if the data provider misbehaves and is subsequently slashed, ensuring economic safety even if data providers act unpredictably.

We will describe this process in more detail in the following sections.
\subsection{Light Client}
\label{subsec:designIns:lightClient}
The light client bootstraps as before, but this time, the active data provider set will include an additional attribute: the ``attributable'' stakes of each data provider, which represent the amount of stakes that have not been assigned to any existing insured requests. Before purchasing the insurance, the light client must determine two key parameters: the coverage amount and its duration.

\paragraph{Coverage amount \vcov.} 
The light client calculates the maximum potential profit that data providers could gain from corrupting the data as the coverage amount \vcov. Then to establish the desired level of security, the light client select the set of data providers, specify the proportion of their stakes needed for the insured request, whose total attributable stakes exceed the coverage amount. 

For instance, a low-value transaction inclusion check might be sufficiently secured by a portion of the stake from a single data provider who has staked 32 ETH. In contrast, a high-value bridge transaction might require backing from 80\% of the total assets staked by multiple data providers. When a light client chooses to rely on a stake from a single data provider, only one signature verification is necessary during the protocol execution while selecting more data providers for the same coverage amount incur higher verification cost. Therefore, it is generally preferable to select a smaller number of data providers as long as their combined stakes meet the necessary threshold.

\paragraph{Coverage duration \tcov.} 
Next, the light client sets a challenge period (\(T_{cp}\)) like before, which determines how long it will wait for alerts from watchers. Unlike in \lceco where the light client would wait for the entire challenge period before accepting the target state,  \lcins provides immediate finality. That is, the light client will accept the target state as soon as it receives responses from data providers, but it will keep listening any potential alert from watchers for $T_{cp}$.

After selecting \(T_{cp}\), the light client can determine the coverage duration \tcov for its insurance. For a single state inclusion check, the light client needs to consider $T_{cp}$ to ensure that any detected misbehavior is covered during the insurance duration. Our protocol also allows the light client to cover multiple checks within the same insurance policy. In this case, setting \tcov must take into account \(T_{cp}\) for different checks and the number of \textit{independent state inclusion checks} denoted by \(n\). The total coverage duration must extend beyond the point when \(T_{cp}\) has elapsed for each of these checks to ensure full coverage of all checks. 




Specifically, the coverage duration for a single transaction (\(n=1\)) needs to be set to cover the total time elapsed from the moment the light client sends the insurance payment transaction to the point when challenge period ends, which includes the sum of the following parameters: 

\begin{itemize}
    \item Blockchain finality time ($T_{fin}$, for the insurance payment to get finalized)
    \item Locally selected challenge period ($T_{cp}^0$, to check the inclusion of the insurance payment transaction)
    \item Another locally selected challenge period ($T_{cp}^1$, to check the inclusion of the current transaction)
    \item Estimated communication delay times($\Delta_{comm}$)
    \item Estimated computation and verification time ($\Delta_{comp}$)
\end{itemize}

Now suppose there are multiple transaction inclusions to check; the only additional delay is that the light client needs to wait for multiple challenge periods to cover all the independent inclusion checks. We denote the $i$-th check has challenge period $T_{cp}^i$. Then the coverage duration is:

\begin{equation}
\label{eq:minInsTime}
    T_{cov} \ge\ T_{fin} + \sum_{i=0}^n T_{cp}^i + \Delta_{comm} + \Delta_{comp}
\end{equation}

From this point, we assume \(n=1\) for simplicity without losing generality. For larger $n$, the same procedure needs to get repeated for each of the state inclusion checks.

\paragraph{Purchasing insurance.}
After selecting the parameters, the light client makes an on-chain call to purchase insurance, specifying the chosen data providers, their stake portions summing to the predetermined threshold \vcov, and the desired coverage duration \tcov. The light client then waits for the blockchain's finality time to pass, and uses \lcins to check the inclusion of the insurance payment transaction. Note that the insurance purchasing transaction might be reverted if there is insufficient available attributable stake. This situation could arise if the combined queries from different light clients exceed the available stake of a specific data provider in a block, compared to the provider’s stake from the previous block. The light client must verify whether the transaction was successful or reverted.

\begin{figure*}
    \centering
    \includegraphics[width=0.9\linewidth]{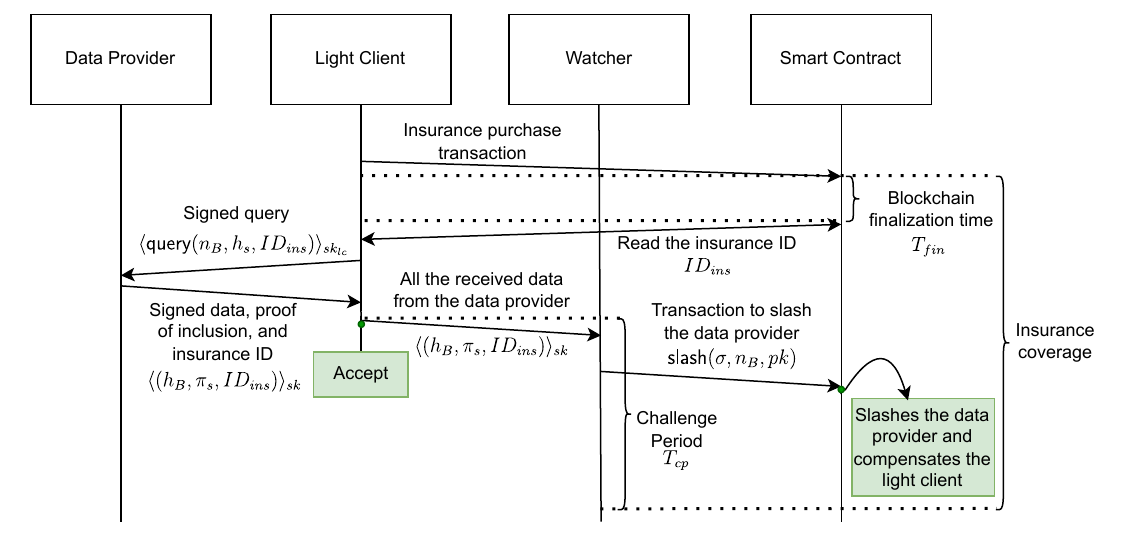}
    \caption{The dispute resolution path for the insured security. We have simplified this example, by supposing the light client has chosen only one data provider. Also, we only show the honest watcher here.}
    \label{fig:ins-dispute}
\end{figure*}

After successfully purchasing insurance, the light client sends queries to the selected data providers. The modification here is that the light client needs to include the insurance ID $ID_{ins}$, assigned by the contract, in the query. The light client signs the query message and sends it to the data providers. After receiving the data and signatures from the providers, it forwards them to the watchers as before. Finally, the light client verifies the signatures and the proof of state inclusion. This time, there is no need to wait for the challenge period, allowing the data to be accepted immediately after signature verification.
If the data providers deliver malicious data, they will be slashed by the watchers. Consequently, the light client will receive the insured amount from the smart contract. The flow of interactions in this design is depicted in Figure \ref{fig:ins-dispute}.

\subsection{Data Providers and Watchers}
\label{subsec:designIns:DPs}
In this variant, watchers function the same as before and there are only a few modifications for data providers. The execution of withdraw request for the data providers differs which we explain in details in section \ref{subsec:designIns:contract}. 
In addition, in response to a query, the data providers also include the insurance ID in the data they sign and send to the light client.

\subsection{On-chain Smart Contract}
\label{subsec:designIns:contract}
Two new functions are essential to the smart contract to facilitate the insurance feature: buying insurance and claiming insurance. Additionally, modifications are necessary for handling withdrawal requests by data providers.

\paragraph{Buying insurance.} 
When this function is invoked, it first validates the inputs (\vcov, \tcov and data providers' public keys) by checking the availability of attributable stakes from the selected data providers. If the stakes are sufficient and available, the contract allocates and locks them for the duration of the insurance, rendering it unavailable for other queries until the insurance expires.
The smart contract assigns a unique ID to each purchased insurance, recording this ID along with the light client’s public key, the involved data providers, and insurance parameters.

\paragraph{Claiming insurance.}
Watchers trigger this function by submitting data signed by data providers that contradicts to on-chain states, including the relevant insurance ID. Upon receipt, the contract verifies the authenticity of the dispute and, if validated, slashes the data providers' stakes. It then allocates the slashed amount to the light client associated with the indicated insurance ID, thus compensating for any breach of security.

\paragraph{Withdrawal request.}
Data providers can exit the system by submitting a withdrawal request. Upon receipt of such a request, the contract changes the provider's status from "active" to "leaving" and stops assigning new insurances to that provider's stake. However, under the insured protocol, portions of the provider’s stake may be locked in active insurances that extend beyond the period of the withdrawal request. The provider must continue to serve these commitments until all associated insurances expire. Only after the expiration of the last active insurance can the provider fully disengage from the protocol, and the withdrawal of their stake can be processed at the end of the subsequent update epoch.

\subsection{Analysis}
\label{subsec:designIns:analysis}
\paragraph{Scalability}
In this design, the capacity of the system is defined as the sum of the total stakes of all rational data providers. They are the ones providing service for the light clients and are incentivized to do so. Since their stake gets attributed to the queries, the maximum fully-covered insured amount can be the sum of the total stakes of all rational data providers. Larger values for state checks cannot be processed by the system, ensuring full coverage. Subsequently, the maximum rate at which the system can insure value and support inclusion checks is limited to this total amount for a single challenge period.

\paragraph{Cost of insurance}
We observe a crucial relationship between the challenge period (\(T_{cp}\)) the light client waits for in \lceco and the cost the light client incurs in \lcins for purchasing insurance. If the light client prefers not to wait for the challenge period, it can opt to pay for insurance instead, thereby bypassing the wait entirely. The cost of this insurance is directly linked to the level of security the light client requires, which in turn determines the challenge period that the light client seeks to avoid. A higher security level necessitates a longer challenge period, meaning that the data providers' stakes need to be locked for a longer duration. Consequently, the light client must pay more for the cost of locking this collateral for the extended period.

To evaluate the cost of insurance for light clients, let's consider a realistic example. Assume we have \(n\) data providers recognized by the smart contract as active for at least one block over the course of a year, which consists of \(B\) blocks. With an average block interval time of 12 seconds on Ethereum, this gives us \(B=2,628,000\) blocks per year.

If a node exits and then rejoins the system, it is counted as two separate data providers. Let \(S_i\) denote the total staked value for the \(i\)-th data provider. In return for staking, data providers earn rewards, representing the cost of their locked stakes. Assume an Annual Percentage Rate (APY) of \(6\%\) for data providers, as determined by market equilibrium.

We define \(u\) as the utilization ratio of the stakes, where utilization refers to the proportion of the stake locked on the contract to hold the corresponding data provider accountable for a specific query. This stake is locked when a user purchases insurance for a designated duration. The parameter \(u\) represents the average portion of the total stake utilized by users at any given time throughout the year:

\begin{align}
    u = \frac{\sum_{t=1}^B u_t}{B} 
\end{align}
where
\begin{align}
    u_t = \frac{\sum_{i\in DP(b_t)} S^{lock}_i}{\sum_{i\in DP(b_t)} S_i}
\end{align}

where \(DP(b_t)\) is the set of data providers with active status on the smart contract in block \(b_t\) and $S_i^{lock}$ represents the locked stake of the $i$-th data provider. Now, we can define the cost of locking a unit of stake for one block, \(c\), as follows:

\begin{align}
    c = \frac{APY}{Bu} 
\end{align}

Light clients determine a coverage period \tcov and an insurance value \vcov. The cost of insurance required to be paid to the data providers is given by 
\begin{align}
    p_{ins} = c \cdot T_{cov} \cdot V_{cov}
\end{align}
where a lower bound for \( T_{cov} \) is determined from \cref{eq:minInsTime}.

Knowing this, for example, to check a single transaction valued at 100 ETH, with a 5-hour challenge time (1500 blocks), and assuming \(u = 0.75\), the cost of insurance would be:

\begin{align}
    p_{ins} = \frac{0.06}{2628000\times 0.75} \times 1500 \times 100 = 0.004566 \text{ ETH}
\end{align}

This calculation does not include the gas cost of the transaction to purchase the insurance, which we will evaluate in the experiments (Section~\ref{sec:eval}).
\section{Evaluation}
\label{sec:eval}
The main objective of our experimental evaluations is to answer the following questions for light clients:
\begin{itemize}
    \item What is the computational overhead?
    \item What is the latency before the light client can confirm the data?
    \item What is the incurred cost?
\end{itemize}

Our light client protocol's smart contract, implemented in Solidity with less than 300 lines of code, is deployed on the Ethereum testnet, Sepolia~\cite{sepolia}, which simulates the consensus mechanism of Ethereum. We developed the backend nodes in TypeScript, including the light client (approximately 350 lines), data providers (approximately 150 lines), and watcher nodes (approximately 200 lines). These nodes communicate over a local network via the HTTP protocol.

In our scenario, we assume that every data provider registered on the smart contract has staked 32 ETH, equivalent to an Ethereum validator's stake. Additionally, we have one light client aiming to verify a transaction valued at less than 32 ETH. Consequently, this light client needs to query only a single data provider. For simplicity, the light client is connected to exactly one honest watcher.

We initially set up two Sepolia full nodes, each on an AWS Lightsail instance with 16 GB RAM, 4 vCPUs, and 500 GB SSD. We use Geth~\cite{geth} as the execution client and Prysm~\cite{prysm} as the consensus client. We then deploy the data provider and watcher nodes on separate instances, configuring their web3 provider to their respective local full nodes. This setup allows the nodes to access the latest blockchain data directly. The light client is deployed on a Lenovo ThinkPad X1 Carbon Laptop, featuring an Intel Core i7 processor, 16 GB RAM, and a 512GB SSD, though the actual hardware requirements are significantly lower.

Our experiments focus on both the economic security protocol \lceco and the insured security protocol \lcins, specifically verifying the finalization of a block hash on the light client. We exclude the time and computation required to verify the Merkle proof of inclusion for its target state, as this is consistent across all light client protocols.

\paragraph{Computation.} The computation required by the light client is consistent in both designs. In the economic design \lceco, the light client begins verification after the challenge period elapses. In the insured design \lcins, verification starts once the inclusion of the insurance payment transaction is confirmed. In both cases, the light client performs one signature verification check on the block hash, taking an order of milliseconds.

\begin{figure}
    \centering
    \includegraphics[width=1\linewidth]{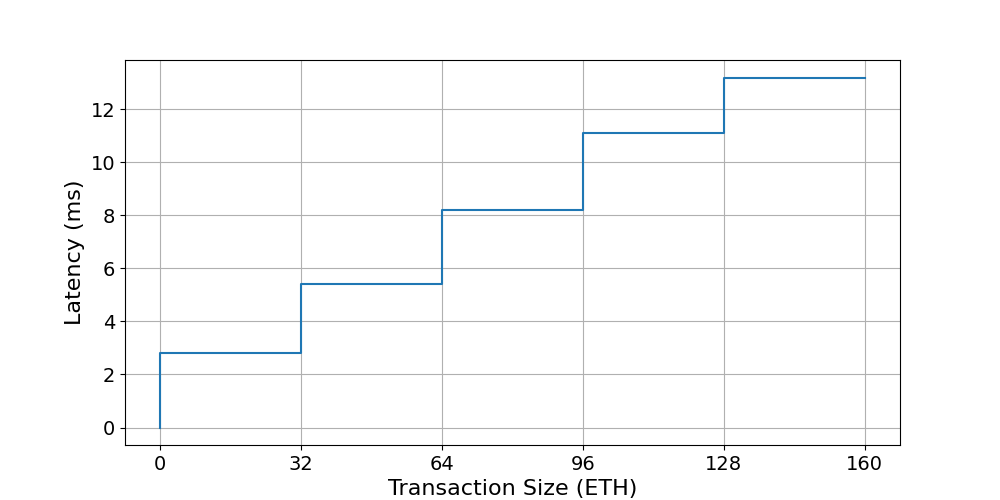}
    \caption{Latency of verifying the data provider signatures for the light client.}
    \label{fig:latvstx}
\end{figure}

\paragraph{Latency.} We varied transaction sizes in our experiments, requiring signatures from 1 to 5 data providers, and measured the latency. As depicted in Figure \ref{fig:latvstx}, the delay is only a few milliseconds. Consequently, in our economic security design, latency is primarily influenced by the challenge period, as computation and communication delays are minimal. The challenge period, configurable by the light client, typically spans several hours but could be reduced to minutes with a fast and reliable watcher network. In contrast, in insured security scenarios, latency depends solely on the time needed to complete signature verifications, which scales linearly with the number of required provider signatures. Given that this process usually takes only milliseconds, we achieve instant confirmation.

\paragraph{Cost.} In order to provide economic security, the light client incurs no additional costs in normal path. However, to provide insured security, the cost for the light client to purchase insurance is divided into two parts: the transaction gas fee and the insurance premium. The transaction gas cost for calling \texttt{buyInsurance()} with inputs from one data provider's address is 200k. Additionally, there is the cost of the insurance itself, as calculated in the previous section for the expected stake return rate of $APY=6\%$. Given the current ETH price of \$3200 and gas cost of 9.377 Gwei, the transaction gas fee amounts to \$6. The insurance premium costs \$1.45, totaling \$7.45 for buying insurance covering 10 ETH (\$32k) for 1500 blocks (approximately 5 hours). We repeat the experiment for different transaction values and different number of data providers. Both the transaction cost and the insurance premium cost increase with the covered value. The results are summarized in Table~\ref{tab:insurance_cost}.

We do not account for any gas costs for an honest watcher, as we assume its honesty and diligence in checking the data. However, in practice, using watcher services may incur costs comparable to a token swap operation on Uniswap \cite{sheng2024proof}. Additionally, the cost incurred by a watcher for submitting a dispute on-chain is compensated from the slashed stake of the data provider implicated in that dispute. 

    

\begin{table*}[]
\centering
\caption{Cost for applying each protocol for checking the inclusion of transactions with values 10 ETH(\$32k), 32 ETH(\$100k), 160 ETH(\$512k), and 320 ETH(\$1M) respectively for 1, 1, 5, and 10 providers (or multiple transactions summing up to these values). Computation is in terms of number of signature checks.}
\label{tab:insurance_cost}
\begin{tabular}{|c|clll|clll|}
\hline
 &
  \multicolumn{4}{c|}{\lceco} &
  \multicolumn{4}{c|}{\lcins} \\ \hline
\multicolumn{1}{|l|}{\textbf{Number of data providers}} &
  \multicolumn{1}{l|}{1} &
  \multicolumn{1}{l|}{1} &
  \multicolumn{1}{l|}{5} &
  10 &
  \multicolumn{1}{l|}{1} &
  \multicolumn{1}{l|}{1} &
  \multicolumn{1}{l|}{5} &
  10 \\ \hline
\textbf{Cost} &
  \multicolumn{1}{c|}{\$0} &
  \multicolumn{1}{l|}{\$0} &
  \multicolumn{1}{l|}{\$0} &
  \$0 &
  \multicolumn{1}{c|}{\$7.45} &
  \multicolumn{1}{l|}{\$10.68} &
  \multicolumn{1}{l|}{\$29.38} &
  \$52.76 \\ \hline
\textbf{Computation} &
  \multicolumn{1}{c|}{1} &
  \multicolumn{1}{l|}{1} &
  \multicolumn{1}{l|}{5} &
  10 &
  \multicolumn{1}{c|}{1} &
  \multicolumn{1}{l|}{1} &
  \multicolumn{1}{l|}{5} &
  10 \\ \hline
\textbf{Latency} &
  \multicolumn{1}{c|}{5 hrs} &
  \multicolumn{1}{l|}{5 hrs} &
  \multicolumn{1}{l|}{5 hrs} &
  5 hrs &
  \multicolumn{1}{c|}{2.8 ms} &
  \multicolumn{1}{l|}{2.8 ms} &
  \multicolumn{1}{l|}{13.2 ms} &
  25.8 ms \\ \hline
\end{tabular}
\end{table*}

\section{Discussion}
\label{sec:discussion}
\subsection{Diverse Staking}
So far, we have assumed that data providers stake or re-stake ETH. This compels light clients to calculate the maximum loss value in ETH, even if their loss is in another token whose value changes rapidly in relation to ETH. This necessitates complex calculations for buying insurance and imposes some risks for the light client due to unpredictable market movements.

To address this, staking can be diversified in terms of the staked token. Data providers can stake other widely used tokens like stablecoins or wrapped Bitcoin, offering more options for light clients to buy insurance. This also enables more data providers to join the system as it allows them to be exposed to the price of their token of choice while staking. However, there may still be light clients using unsupported tokens or having more complex loss functions. 

\subsection{Delegation}
As mentioned before, the preference is to have fewer data providers with larger amounts of stake to expedite the verification process for the light client. However, to allow data providers with less valuable assets to join the protocol, delegation can be employed. This involves smaller data providers pooling their stake and delegating their data provision duties to a single node. This is not a new concept and has been utilized in various blockchain contexts.

In delegation, the reward is distributed among all stakers proportionally to their staked value, with some additional reward going to the operator node. Delegation is also beneficial for users who are unfamiliar with or lack access to the necessary resources for running a data provider node but have assets they wish to stake.

\subsection{Proposed Block Guarantee}
So far, we have only discussed the guarantee that a block has been finalized. On Ethereum, every proposed block will eventually become either an uncle block or finalized. Another useful guarantee that can be supported by this design is to ensure whether a block has been proposed in the blockchain. It might not yet be finalized, but since finalization in Ethereum takes roughly 13 minutes, to avoid this wait, a light client can use this new type of guarantee to access the data faster. The block might not get finalized later, but that happens with a very low probability. Such a guarantee is useful for low-value transactions that require low latency.

To add this feature to the system, the light client needs to add a flag to the query it sends to the data provider determining which type of guarantee it wants. Then, the data providers also need to include the same flag in their signed data so that when the watchers provide it on-chain, the smart contract knows which guarantee to check this data against.

The data provider that signs the block data will get slashed only if the signed block header was not ever proposed in the blockchain. To support this guarantee, we need to modify the design slightly. For watchers to be able to provide a proof to the smart contract that a data provider has signed incorrect data, they have to prove to the contract that a certain block has not been proposed in the blockchain at all. Meaning, it is neither finalized nor an uncle block. This can be done optimistically, requiring the data providers to provide inclusion proofs if disputed. The inclusion proofs can be in form of zero-knowledge proofs to reduce the cost of on-chain verification. Moreover, the watchers need to stake some assets in the system, with each watcher staking an amount greater than the gas fee of the call to provide the necessary proof in case of a dispute. This way, the watchers do not have incentive to dispute falsely.

Data providers need to submit on-chain inclusion proofs in case of a dispute. If the data provider has not misbehaved, there are two possibilities: either the block has become finalized, or it has become an uncle block. The proof for the latter case is to check the inclusion of the block in the state root of the latest block header. For the former, in addition to the inclusion proof, finality check is also needed. Here, we leverage the fact that uncle blocks are included in the state root commitment, as well as the finalized blocks.

\subsection{Cost and Fee Management}
The fee that the light clients pay for the insurance goes for the compensation of the data providers. Data providers get compensated proportional to the time and amount of the insurance they provide for the light clients. 

Watchers also get compensated for the services they provide. They get a percentage of the slashed stake whenever they slash a data provider successfully. This way they stay incentivized to actively watch the data they receive from the light clients.

\subsection{Data Availability}
Since data providers operate full nodes and maintain the complete blockchain data, they are capable of supporting not only consensus verification services but also data availability checks. Previous studies~\cite{albassam2019fraud,yu2020coded,sheng2021aced} have designed efficient data structures for light nodes to verify data availability. Overall, there are two basic models: a ``pull'' model in which light clients randomly sample data from full nodes, and a ``push'' model in which block producer disperses different data chunks to data providers. The ``pull'' model can be directly adopted here under the same security assumption. Staked data providers are responsible for answering sampling requests from the light client. The light client then forwards data samples to an honest validator node until the validator can either reconstruct the block or provide an incorrect-coding proof. The fraud proof can be submitted on-chain to penalize malicious data providers. A related protocol~\cite{tas2023cryptoeconomic} also examines the cryptoeconomic security for the data availability committee. Our light client protocol further offers insights into extending cryptoeconomic security to insured security, utilizing penalties to compensate for losses caused by data availability attacks.
\section{Conclusion}
\label{sec:conclusion}
In this paper, we have formalized the cryptoeconomic security for light clients and introduced programmable security options tailored to their needs. We presented two economically robust designs for a light client, focusing on Ethereum PoS. The first design is more cost-effective for the light client but introduces a higher latency. The second design allows the light client to trust the data almost instantaneously, albeit for a small fee. Importantly, in this design, the light client is compensated if the provided data proves incorrect. This work introduces the first economically safe light client protocol, serving as a pivotal component for various applications that need to verify transaction inclusions to secure the finalization of their payments. These received payments might be in exchange for services or goods provided to their counterparts, or can be the requests in applications like bridges that mint value upon verifying the finalization of a payment. In all cases, the applications can enjoy almost instant verification while being insured for the value of their payments.


\bibliographystyle{ACM-Reference-Format}
\bibliography{bib}


\end{document}